\documentclass[journal = apchd5, manuscript = article, layout =  traditional]{achemso}

\usepackage{chemformula} 
\usepackage[version=3]{mhchem}
\usepackage[T1]{fontenc} 
\usepackage{graphicx}
\usepackage{subcaption} 
\usepackage{multicol}
\usepackage{upgreek}
\newcommand{\onlinecite}[1]{\hspace{-1 ex} \nocite{#1}\citenum{#1}}

\author{Evgeniia Slivina}
\affiliation{Meyer Burger Research AG, Rouges-Terres 61, Hauterive 2068, Switzerland }
\alsoaffiliation{Institute of Nanotechnology, Karlsruhe Institute of Technology, Hermann-von-Helmholtz-Platz 1, Eggenstein-Leopoldshafen 76344, Germany}
\email{evgeniia.slivina@kit.edu}
\author{Aimi Abass}
\affiliation{Institute of Nanotechnology, Karlsruhe Institute of Technology, Hermann-von-Helmholtz-Platz 1, Eggenstein-Leopoldshafen 76344, Germany}
\author{Derk B$\ddot{\text{a}}$tzner}
\affiliation{Meyer Burger Research AG, Rouges-Terres 61, Hauterive 2068, Switzerland }
\author{Benjamin Strahm}
\affiliation{Meyer Burger Research AG, Rouges-Terres 61, Hauterive 2068, Switzerland }
\author{Carsten Rockstuhl}
\altaffiliation{Institute for Theoretical Solid State Physics, Karlsruhe Institute of Technology, Wolfgang-Gaede-Str. 1, Karlsruhe 76131, Germany}
\author{Ivan Fernandez-Corbaton}
\affiliation{Institute of Nanotechnology, Karlsruhe Institute of Technology, Hermann-von-Helmholtz-Platz 1, Eggenstein-Leopoldshafen 76344, Germany}

\title[An \textsf{achemso} demo]
  {Insights into backscattering suppression in solar cells from the helicity preservation point of view}

\abbreviations{IR,NMR,UV}

\begin{document}


\begin{abstract}
	We show that the anti-reflection performance of nano-particle arrays on top of solar cell stacks is related to two conditions: a high enough degree of discrete rotational symmetry of the array and the ability of the system to suppress cross-talk between the two handednesses (helicities) of the electromagnetic field upon light-matter interaction. For particle-lattice systems with high enough degree of discrete rotational symmetry $2\pi/n$ for $n\ge3$), our numerical studies link the suppression of backscattering to the ability of the system to avoid the mixing between the two helicity components of the incident field. In an exemplary design, we optimize an array of TiO$_2$ disks placed on top of a flat heterojunction solar cell stack and obtain a three-fold reduction of the current loss due to reflection with respect to an optimized flat reference. We numerically analyze the helicity preservation properties of the system, and also show that a hexagonal array lattice, featuring a higher degree of discrete rotational symmetry, can improve over the anti-reflection performance of a square lattice. Importantly, the disks are introduced in an electrically decoupled manner such that the passivation and electric properties of the device are not disturbed. 
\end{abstract}

\section{Introduction}
Minimal reflection is an obvious design goal in solar cell technology which attracts much research attention. Different approaches to anti-reflection (AR) range from chemical texturing of the silicon waver \citep{papet2006pyramidal,kim2008texturing} to sophisticated AR coatings \citep{lien2006tri, du2010hollow, shimomura2010layer, barshilia2012nanometric, liu2012broadband} and plasmonic structures \citep{catchpole2008plasmonic,ferry2008plasmonic,green2012harnessing}. Recently, the use of arrays of dielectric nano-structures is being investigated as a possible avenue to improve AR properties of solar cells \citep{ferry2011modeling,spinelli2012broadband,spinelli2014light,kim2014silicon,baryshnikova2016plasmonic,cordaro2019antireflection}. Due to the low profiles of the patterning nanostructures, this approach is suitable for ultra thin film solar cells. Clearly, understanding the underlying physical principles behind backscattering minimization is relevant for the AR aspect of solar cell design.

Much of the nanophotonics research on backscattering minimization stems from the 1983 article of Kerker {\em et al.} \citep{kerker1983electromagnetic}. This early work showed that a sphere whose relative electric permittivity and magnetic permeability are equal exhibits zero backscattering under plane wave illumination, i.e., there is no energy in the specular back reflection direction, independently of the polarization of the illuminating plane wave. Since then, the theoretical and experimental works on zero backscattering have been numerous, see e.g. Refs. \onlinecite{Lindell2009,Nieto2011,zambrana2013duality,fernandez2013forward,Coenen2014,Alaee2015b,Geffrin2012,Person2013,Abdelrahman2017}. On the theoretical side, the relationship between electromagnetic duality symmetry and zero backscattering \cite{Lindell2009,zambrana2013duality,FerCor2012p} has provided a new point of view on Kerker's result by connecting the backscattering suppression to a fundamental symmetry in electromagnetism. A system is symmetric under duality transformations \cite[Sec. 6.11]{Jackson1998} if and only if its electric and magnetic responses to incident radiation are equivalent. This equivalence connects directly to Kerker's $\epsilon=\mu$ spheres. In much the same way that translational symmetry preserves linear momentum, and cylindrical symmetry preserves angular momentum, duality symmetry preserves the helicity (handedness) of any incident illumination (see e.g. Ref. \onlinecite{FerCor2012p}). That is, the interaction of light with a dual symmetric object does not couple the left and right handed components of the electromagnetic field. In terms of requirements, duality symmetry of the system is not enough to ensure zero backscattering, but duality plus cylindrical symmetry is \cite{zambrana2013duality,FerCor2012p}. The two conditions together fully explain the effect reported by Kerker {\em et al.} in $\epsilon=\mu$ spheres. In terms of practical application in solar cells, the $\epsilon=\mu$ condition is an obvious roadblock, and full cylindrical symmetry is also not achievable by regular arrays of nanoparticles. However, the work in Ref. \onlinecite{fernandez2013forward} established that, for a general system, a discrete rotational symmetry $2\pi/n$ for $n\ge 3$ and helicity preservation upon illumination along the symmetry axis are sufficient conditions for zero back-scattering \cite{fernandez2013forward}. This is a general result whose derivation does not depend on the details of the system, only on its symmetries. The importance of the result in the context of designing regular arrays of dielectric nanostructures for AR applications is twofold. First, it establishes which which kind of array lattices are appropriate. Second, and crucially, it relaxes the duality symmetry condition, which implies helicity preservation for {\em all} illumination directions, into helicity preservation upon normal incidence. While exact duality symmetry in macroscopic electrodynamics requires $\epsilon=\mu$ materials, helicity preservation under particular illumination conditions can be achieved by geometrical optimization for systems where $\epsilon\neq\mu=1$. The geometry of the structures needs to result in an equal electric and magnetic response of the system under the prescribed illumination. For a dielectric disk, the tuning of its aspect ratio achieves the desired effect for on-axis illumination at a particular frequency by aligning the electric and magnetic dipolar resonances of the disk \cite{Evlyukhin2011,Staude2013,Chong2016}. This makes dielectric disks a suitable candidate for the constituents of the AR arrays. Indeed, square and hexagonal arrays of optimized silicon cylinders showing almost zero reflection have been reported in the solar cell literature \citep{spinelli2014light,kim2014silicon}, and also in other applications where high transmission is required \cite{Staude2013,Chong2016}. 

In this article, we use full-wave numerical calculations to show that helicity preservation and a high enough degree of discrete rotational symmetry are the relevant requirements for the design of nanoparticle arrays for reflection minimization in solar cells. We first use hypothetical materials with $\mu\neq1$ to illustrate the effects of both conditions on the AR properties of the array. For particle-lattice combinations with enough discrete rotational symmetry, our numerical studies link the suppression of backscattering to the ability of the system as a whole to preserve the helicity (handedness) of the incident light. Then, we consider a realistic heterojunction (HJT) layer stack and design an array of TiO$_2$ disks whose current loss due to reflection is improved three times with respect to an optimized flat reference. We analyze the design in terms of helicity preservation and consider both square and hexagonal lattices, showing the advantages of the latter. Importantly, the TiO$_2$ disks are placed on the top face of the HJT stack in an electrically decoupled manner, which avoids the passivation difficulties brought about by placing dielectric nano-scatterers directly on top of the Si wafer. The low absorption of TiO$_2$ in the spectral region relevant for solar cells is one of the reasons for the choice of this material.

It is important to note that the generality of the helicity preservation plus discrete rotational symmetry conditions make them useful for understanding several apparently different techniques for AR improvement by means of the same physical principles. For example, in standard silicon based solar cells, improvement of the AR properties is often achieved by means of the chemical texturing of a silicon wafer \citep{papet2006pyramidal,kim2008texturing}. In this way, micron-sized pyramids can be formed on the front and/or back surfaces of the wafer. Another example are graded index multilayers \citep{lien2006tri,shimomura2010layer}. Both techniques can be understood with the symmetry principles described above. In a system composed by several different materials indexed by $r=1,2,\ldots$, including vacuum, the condition $\epsilon_r=\mu_r$ is sufficient for duality and hence for helicity preservation\cite{FerCor2012p}, and the amount of helicity conversion can be tied to the magnitude of the gradient of the impedance across the system $\nabla \sqrt{\frac{\mu\left(\mathbf{r}\right)}{\epsilon\left(\mathbf{r}\right)}}$ [see Eqs. (2.12)-(2.14) in Ref. \onlinecite{Birula1996}]. In this context, the good AR performance of the pyramids can be explained because they provide a low impedance gradient transition between air and silicon when advancing from their tips to their bases. With respect to the additionally required degree of rotational symmetry, randomly positioned pyramids achieve an effective cylindrical symmetry, while perfectly ordered pyramids have $C_4$ symmetry. Both cases meet the requirement. This explanation can be compared with the common argument which explains the backscattering reduction by multiple reflections on the facets of the pyramids. This argument typically involves the ray-optics approximation, which is not really appropriate for sizes of the pyramids in the order of a few wavelengths. The symmetry based explanation does not depend on any such approximation. In graded index multilayers, the minimization of the impedance gradient across interfaces is obvious, and, as flat systems, feature cylindrical ($C_{\infty}$) symmetry.  

The rest of the paper is structured as follows. In Sec. \ref{sec:results}, we start by using hypothetical materials with $\mu_r\neq1$ in order to show the influence that the degree of rotational symmetry and helicity preservation have on the AR properties of a multilayer stack decorated with an array of nanoparticles. We then consider natural materials ($\mu_r=1$), and optimize the nanoparticle array on top of a HJT solar cell layer stack. Later, we show how the helicity preservation properties of a single nanoparticle response correlate with the AR properties of an array of such particles. We finish with a more detailed analysis of the effect of the lattice symmetry.

All numerical full-wave simulations have been done with the finite-element based Maxwell solver $JCMsuite$. The Methods section contains details about the calculated quantities. The refractive index information can be found in the Supp. Mat. 

\section{Results and discussion\label{sec:results}}
A few introductory definitions are in order before we start. The helicity operator $\Lambda$ is defined as the projection of the angular momentum operator vector onto the direction of the linear momentum operator vector, which, for spatially dependent monochromatic fields is simply the curl operator divided by the wave number:
\begin{equation}
	\label{eq:heldef}
	\Lambda=\frac{\mathbf{J}\cdot\mathbf{P}}{|\mathbf{P}|}\equiv \frac{\nabla\times}{k}.
\end{equation}
Electromagnetic duality is a continuous transformation that rotates electric and magnetic fields onto each other (see \cite[Eq. 6.151]{Jackson1998}):
\begin{equation}
	\begin{split}
		\mathbf{\bar{E}}&=\mathbf{E}\cos\theta + Z\mathbf{H}\sin\theta,\\
		Z\mathbf{\bar{H}}&=Z\mathbf{H}\cos\theta - \mathbf{E}\sin\theta,
	\end{split}
\end{equation}
	where $Z$ is the medium impedance. Helicity and duality are related like angular momentum and rotations or linear momentum and translations: helicity is the generator of the duality transformations. Correspondingly, they are tied by a conservation law. In the same way that a cylindrically symmetric system preserves angular momentum or a translationally symmetric system preserves linear momentum, an electromagnetically dual system preserves helicity. That is, the light-matter interaction with a dual symmetric object does not couple the left and right handed components of the electromagnetic field, which are defined as $\sqrt{2}\mathbf{G}_\pm=\mathbf{E}\pm i Z\mathbf{H}$: the Riemann-Silberstein vectors \cite{Birula1996,Birula2013}, which are the eigenstates of the helicity operator with eigenvalues $\pm 1$. A general $\mathbf{G}_+(\mathbf{G}_-)$ field can always be decomposed into a linear combination of left(right) handed polarized plane waves. In the Methods section, Eq. (\ref{eq:hel}) defines the left and right handed polarization vectors as a function of the wavevector direction. For example, for a wavevector along the $\hat{\mathbf{z}}$ direction, $\left(-\hat{\mathbf{x}}-i\hat{\mathbf{y}}\right)/\sqrt{2}$ corresponds to the left handed polarization or helicity +1, and $\left(\hat{\mathbf{x}}-i\hat{\mathbf{y}}\right)/\sqrt{2}$ corresponds to the right handed polarization or helicity -1. For a wavevector along $-\hat{\mathbf{z}}$, the left (+1) and right (-1) handed polarization vectors are $\left(\hat{\mathbf{x}}-i\hat{\mathbf{y}}\right)/\sqrt{2}$ and $\left(-\hat{\mathbf{x}}-i\hat{\mathbf{y}}\right)/\sqrt{2}$, respectively. Note that the +1 helicity vector for $\hat{\mathbf{z}}$ is identical to the -1 helicity vector for $-\hat{\mathbf{z}}$, and viceversa. This illustrates well the meaning of helicity in Eq. (\ref{eq:heldef}): The same sense of rotation together with an opposite linear momentum direction results in an opposite handedness. With this in mind, the main idea behind the sufficient conditions for zero-backscattering derived in Ref. \onlinecite{fernandez2013forward} can be summarized as follows. When a plane wave with definite polarization handedness impinges on a system with a high enough degree of discrete rotational symmetry ($C_{n\ge 3}$), the light transmitted and reflected along the symmetry axis must have the same angular momentum as the illumination. This means that the reflected plane wave must be of changed helicity since, while the angular momentum is the same, the linear momentum is opposite to the one of the incident plane wave. Therefore, if, besides having the $C_{n\ge 3}$ symmetry, the system forbids helicity changes upon normal incidence, the back-scattering will be zero. Since the same argument works for incident plane waves of both helicities, and an arbitrarily polarized plane wave can always be written as the weighted sum of the two helicities, it follows that the system will exhibit zero-backscattering independently of the incident polarization.

While interaction with perfectly dual symmetric systems implies exactly zero coupling between the two helicity components of the incident light, realistic designs will introduce some degree of coupling. For our purposes in this article, it is important to have a measure of such degree of coupling. In our simulations, the illumination is always a plane wave of helicity -1 with a fixed intensity. A continuous measure of the helicity preservation performance, or the degree of helicity preservation, can be then obtained by dividing the total outgoing power of changed helicity by the incident power (see Sec. \ref{sec:xixa}). We use this normalized measure throughout the article.
 
We now start by using hypothetical materials with $\epsilon_r=\mu_r$ to illustrate the influence that the degree of rotational symmetry and helicity preservation have on the AR properties of the system. Figure~\ref{fig01}(a) shows the wavelength dependent reflectance for two systems with rectangular ($C_2$) and square ($C_4$) symmetries, respectively. The different degrees of rotational symmetry are achieved by patterning the top of the same base system with 2D arrays of disks arranged in different lattices: rectangular and square. The ratio of the disks' surface area to the unit cell area of the lattice is kept the same in both arrays. Except for the $\epsilon_r=\mu_r$ materials, the sequence of layers corresponds to an HJT solar cell stack. The illumination is a circularly polarized plane wave whose momentum is normal to the plane of the layer stack. The periodicity of the lattices is always smaller than the free space wavelength, so only the 0$^\text{th}$ diffraction order is present. The results in Fig. \ref{fig01}(a) show that, indeed, the system needs to have a high enough degree of rotational symmetry in order to achieve zero-backscattering: while disks arranged in a square lattice yield zero backscattering, a rectangular lattice does not, despite the perfect electromagnetic (EM) duality of the system. 
\begin{figure}[h]
\centering
  \includegraphics[width=\textwidth]{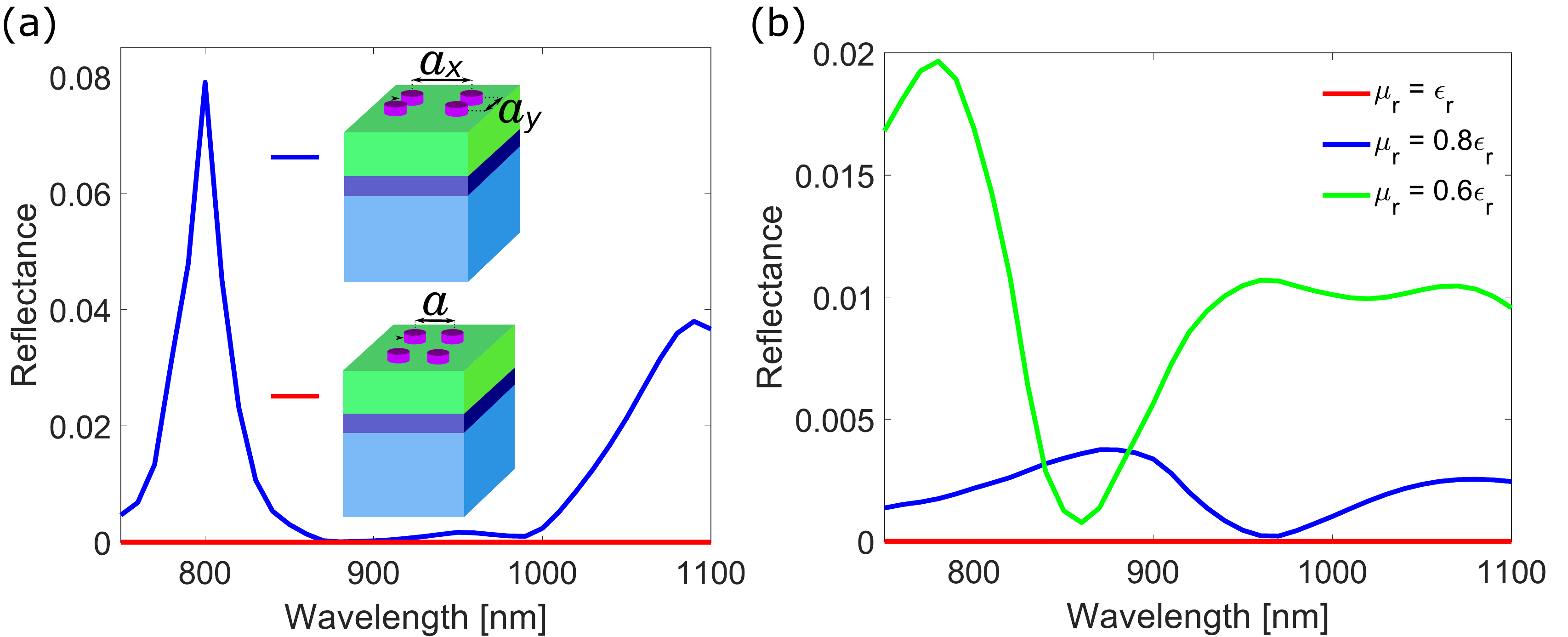}
	\caption{(a) Reflectance from dual-symmetric systems with rectangular ($C_2$) and square ($C_4$) rotational symmetries under normal incidence. The systems are made of hypothetical materials with electric permittivities $\epsilon_r$ equal to those of natural materials and whose magnetic permeabilities are set to $\mu_r=\epsilon_r$ to achieve perfect duality symmetry. The natural materials and geometry are: a semi-infinite c-Si substrate, an absorbing 300nm thick film of cSi, a 8nm thick layer of passivating a-Si (intrinsic and p-doped), an ITO layer of 50nm thickness, and TiO$_{2}$ disks with a height of 100nm  and a diameter of 300nm. The $C_{4}$ unit cell parameter is $a$=500nm. The $C_{2}$ unit cell parameters are $a_{x}$=806.5nm and $a_{y}$=310nm. The unit cells of the systems are schematically shown in the inset. (b) Reflectance from the system with square ($C_4$) rotational symmetry. Colors correspond to different ratios of relative permeability with respect to relative permittivity.} \label{fig01}
\end{figure} 
We now ask the question of how the violation of perfect helicity preservation affects the reflection when the rotational symmetry is sufficiently high. The answer is provided in Fig. \ref{fig01}(b), where the reflectance from a C4 system with different $\epsilon_r/\mu_r$ ratios is shown. The results clearly show that the AR properties gets worse as the materials become less dual symmetric. Figures \label{fig01}(a,b) show the design requirements for an anti-reflection coating: a high enough degree of discrete rotational symmetry and helicity preservation.

While the first requirement is readily met, the lack of $\epsilon=\mu$ materials in the relevant spectral region prevents us from meeting the helicity preservation requirement in this way. Fortunately, helicity preservation under particular illumination conditions can still be achieved for systems where $\epsilon\neq\mu=1$ by geometrical optimization. The geometry needs to result in an equal electric and magnetic response of the system under the prescribed illumination. For dielectric disks, the tuning of their aspect ratios achieves the desired effect for on-axis illumination at a particular frequency \cite{Evlyukhin2011,Staude2013,Chong2016}. We note that this is not equivalent to duality symmetry, because duality would imply that the disk preserves helicity for {\em all} illumination directions. Electromagnetically small dielectric spheres can actually be design to achieve an omnidirectionally high degree of helicity preservation at a particular frequency \cite{zambrana2013duality}. Contrary to the disks, though, this does not happen at resonant frequencies. The degree of helicity preservation in designs based on $\epsilon\neq\mu=1$ materials depends on the frequency and, often, on the illumination direction.

\begin{figure}[h]
\centering
  \includegraphics[width=\textwidth]{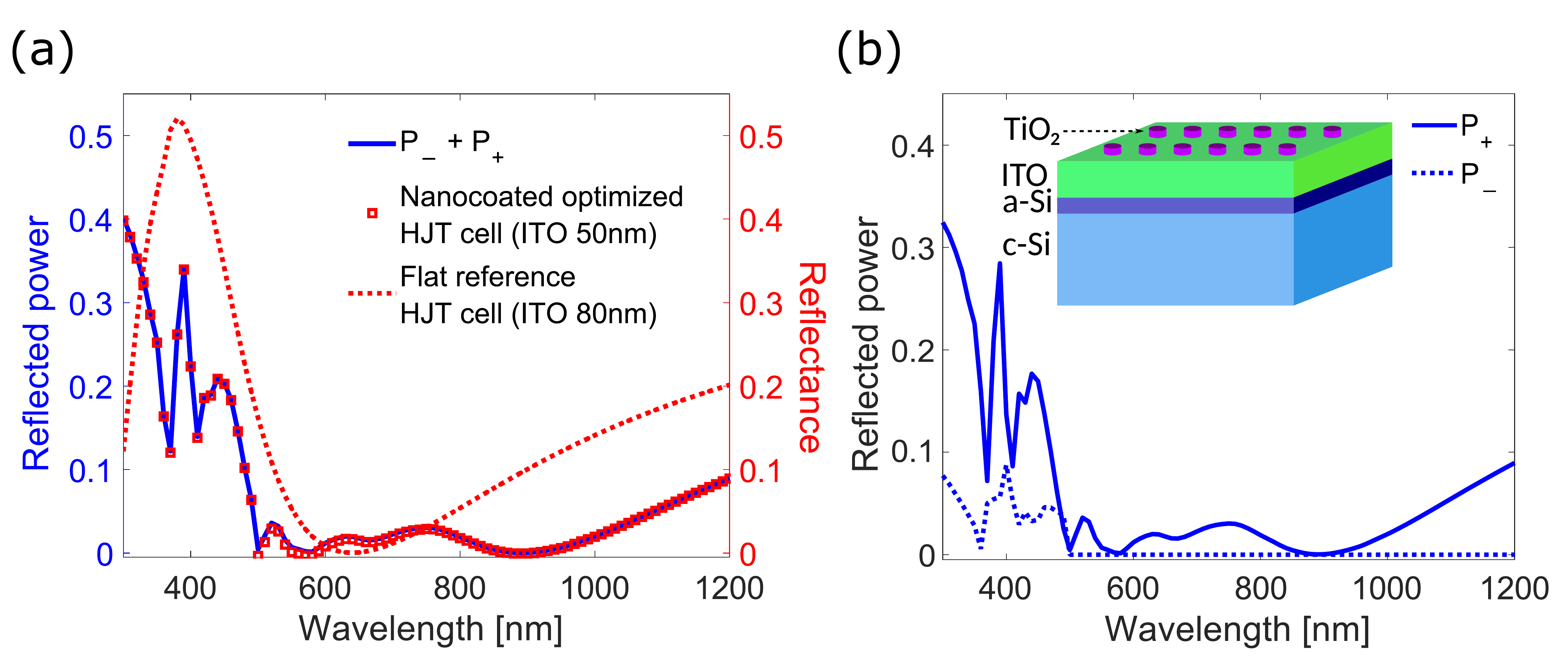}
  \caption{(a) Reflectance versus wavelength for a standard flat HJT solar cell with optimized ITO thickness and for an optimized nano-coated structure. The incident field is a normally incident plane wave with right handed polarization (helicity -1). (b) Normalized reflected powers of EM fields of both helicities $\text{P}_{+}$ and $\text{P}_{-}$ of the square array comprised of TiO$_{2}$ disks placed on the top surface of HJT solar cell. The system is schematically shown in the inset; lattice constant $a =$ 500nm. The layers comprising the substrate are: a semi-infinite c-Si substrate, an absorbing 300nm thick film of cSi, a 8nm thick layer of passivating a-Si (intrinsic and p-doped), an ITO layer of 50nm thickness. The TiO$_{2}$ disks have a height of 100nm and a diameter of 300nm..}  \label{fig2}
\end{figure}

We now consider the C4 system in Fig. \ref{fig01}, this time with $\mu=1$ materials (see the Supp. Mat. for information on the refractive indexes that we use). The system is now representative of an nanocoated HJT solar cell. Figure \ref{fig2}(a) shows the reflectance of the optimized system compared to an optimized flat reference. The optimization target was to minimize the current loss due to reflection upon illumination with a normally incident plane wave of right handed polarization. The height of individual disks, lattice constant, and the thickness of the ITO layer were varied in 10nm steps; the radii of the disks were changed in 25 nm steps. The optimization was done by fixing three out of the four parameters and calculating the current loss from reflectance per unit area as a function of the fourth parameter. After finding the optimal value, the parameter was fixed and another parameter was then varied. Through such iterations the set of parameters corresponding to the lowest possible current loss due to reflection was determined. The optimum is obtained with a configuration of a 500nm lattice constant, ITO thickness of 50nm, and disks of height equal to 100nm and diameter equal to 300nm. For an incident spectrum according to air mass (AM) 1.5, the optimal performance corresponds to a current loss per unit area of 1.7mA/cm$^2$ as calculated with the procedure described in the Methods section. The results are robust to change of the parameters' values, including both disk dimensions and ITO layer thickness, up to 20$\%$. The changes in height of the disks and thickness of the ITO layer lead to slight shifts of the reflectance minima. The variation of the disk radius shifts the minima more significantly whilst keeping the integrated reflectance over the whole frequency range almost unchanged. The 1.7mA/cm$^2$ performance number can be compared to the flat reference, which is identical to the nanocoated one, except that the disks are removed and the ITO layer is re-optimized in 10nm steps for minimal current loss. Such an optimized reference cell achieves a loss of 5.1mA/cm$^2$ for an ITO thickness of 80nm. The choice of a flat reference allows the direct assessment of the gain due to the disk array. Additionally, it can also be argued that the comparison with the flat reference is appropriate in the following sense. The 100nm height of the array would allow its use in extremely thin solar cells, while traditional methods like pyramid etching become impractical beyond thicknesses comparable to the height of the etched elements \cite{brongersma2014light,tan2017small}.

Figure \ref{fig2}(b) shows the (normalized) total power that is reflected back into the air hemisphere for each helicity. To analyze the helicity content, the reflected field is decomposed into plane waves corresponding to negative and positive helicities (see Methods). The qualitative behavior observed in Fig. \ref{fig2}(b) is the expected one. For wavelengths above 500nm where the $C_{4}$-symmetric array is sub-wavelength, only the 0$^\text{th}$ diffraction order is allowed which means that only the specular reflection is allowed. Then, as expected\ and, indeed, seen in the figure for wavelengths longer than 500nm, the $C_4$ symmetry of the array prevents the reflected light from having a component with the same helicity as the illumination \cite{fernandez2013forward}. All the reflected power is of changed helicity. In this regime, {\em the reflection suppression performance is equivalent to the helicity preservation performance}. On the other hand, for wavelengths shorter than 500nm, higher diffraction orders are present. The light scattered back into air is of mixed helicity. This is also expected: the forced helicity flip on backscattering off a $C_{n>2}$ symmetric system applies when the momenta of the incident and reflected plane waves are aligned with the symmetry axis \cite{fernandez2013forward}. This is not met by higher diffraction orders.

\begin{figure}[htbp]
\centering
  \includegraphics[width=\textwidth]{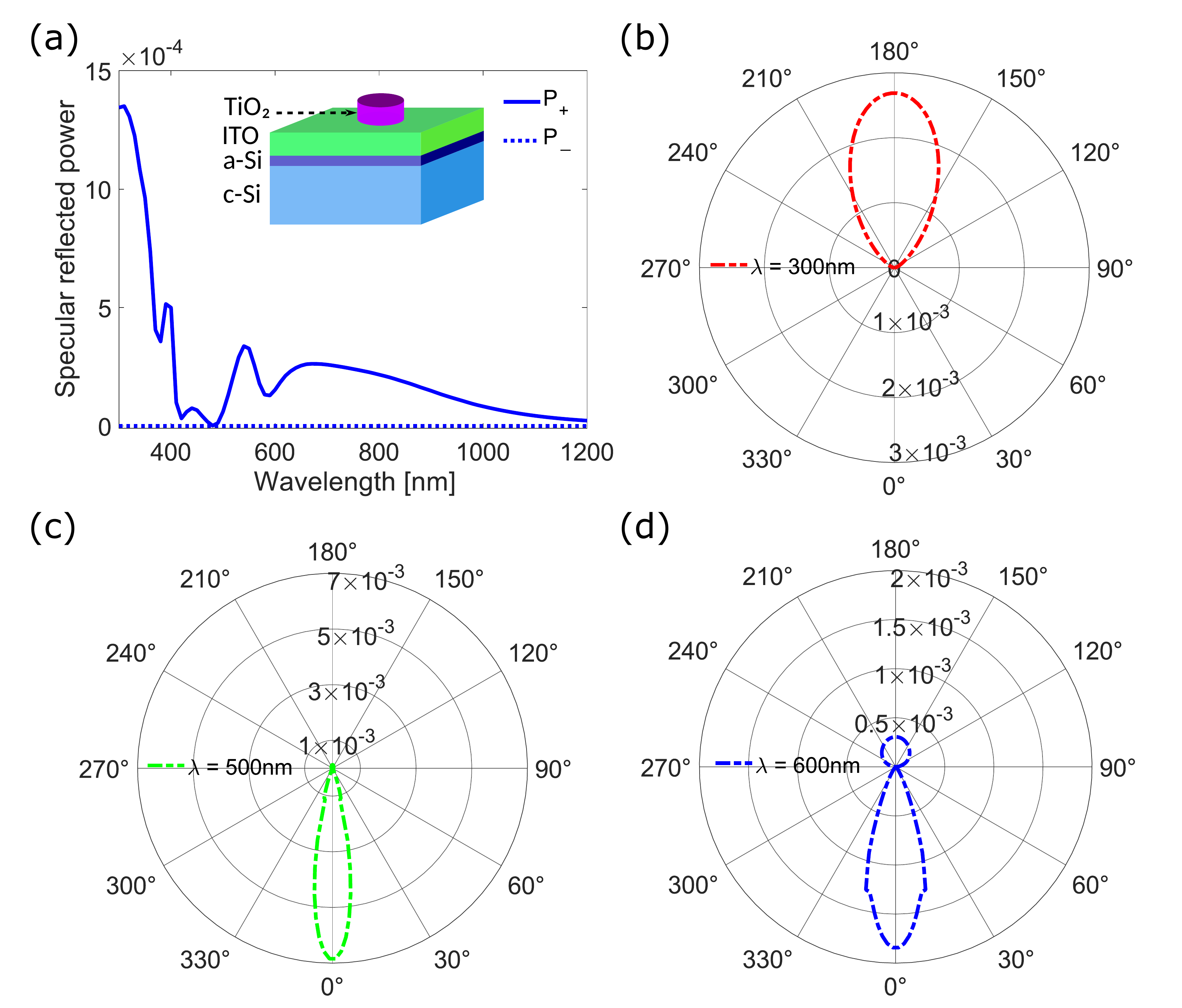}
	\caption{(a) Normalized specularly reflected powers of EM fields of both helicities $\text{P}_{+}$ and $\text{P}_{-}$ for an isolated nano-disk on top of the HJT solar cell layer stack under normal incidence of illumination with a right-handed polarized plane wave (helicity -1). The system is schematically shown in the inset and has the same parameters as in Fig. \ref{fig2} except that we consider here an isolated disk and not an array. (b)-(d) Polar plots of the far field intensities at wavelengths $\lambda$ = 300, 500, and 600nm, respectively. The far field intensities are computed in a plane that contains the cylinder axis of the disk.} \label{fig1}
\end{figure}

In order to get more insight into the results of the optimization, let us now analyze the helicity preserving properties of the selected TiO$_2$ disks in isolation. We consider an isolated nano-disk on top of the layer stack system illuminated by a normally incident plane wave of negative helicity. Figure \ref{fig1}(a) shows the results for the specularly reflected powers corresponding to the two helicities. We observe that only the values of changed helicity $\text{P}_{+}$ are nonzero. As previously discussed, this is expected from the $C_\infty$ cylindrical symmetry of the system, independently of its helicity preservation properties. The system achieves practically zero backscattering at $\lambda$ = 500nm, where $\text{P}_{+}\approx 0$. At this frequency, the system as a whole is approaching perfect helicity preservation. Figures \ref{fig1}(b)-(d) show the far field intensity polar plots for three selected wavelengths. For $\lambda$ = 300nm and  $\lambda$ = 600nm, light is scattered back into air. At $\lambda$ = 500nm, practically all the scattering is towards the forward hemisphere. Being the result of a geometrical optimization, the degree of helicity preservation in designs based on $\epsilon\neq\mu$ materials depends on the frequency and, for disks, also in the illumination direction. With the current designs, the degree of helicity preservation will degrade as the frequency deviates from the optimal one, causing an increase of the back-scattering. We can then ask the question of whether the degree of discrete rotational symmetry can be used to better cope with the currently unavoidable helicity change. 

\begin{figure}[h]
    \centering
    \includegraphics[width=\textwidth]{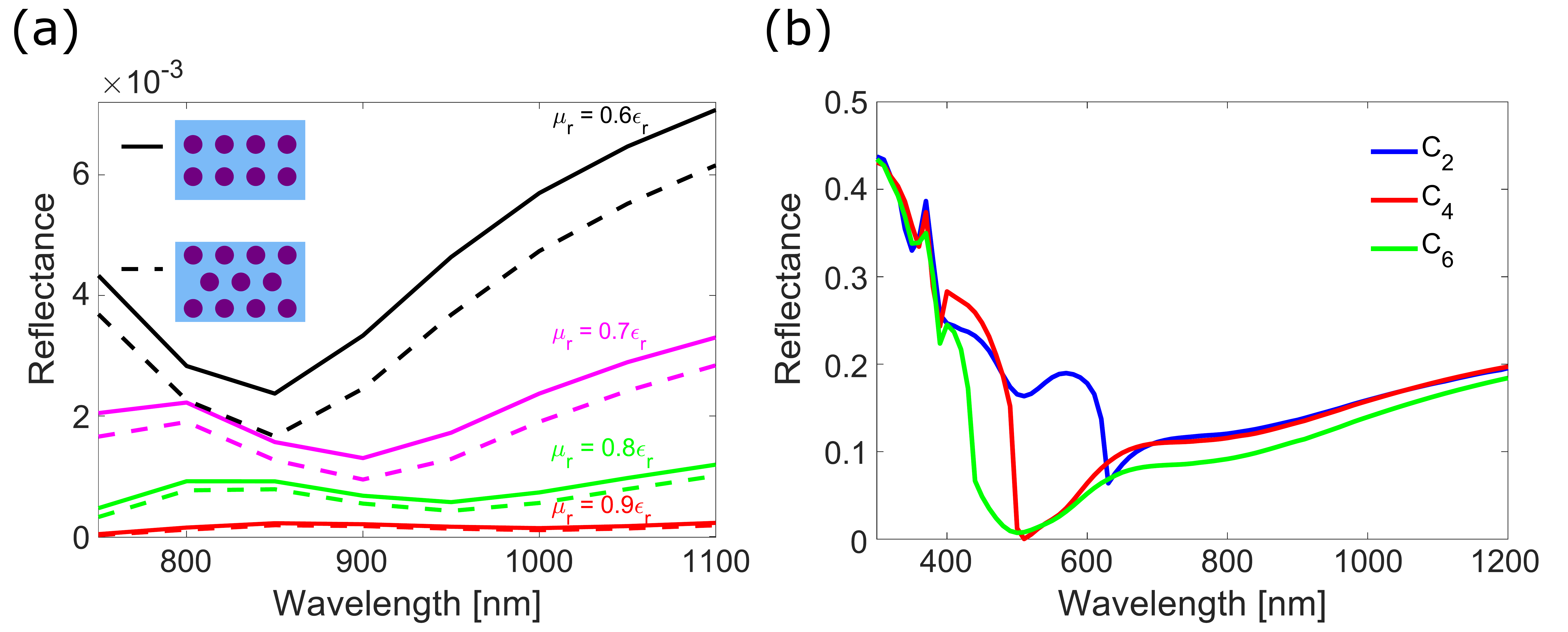}
	\caption{(a) Reflectance from a nano-coated substrate for $C_4$ (bold lines) and $C_6$ (dashed lines) lattices under normal incidence. The electric permittivity of the substrate and disks correspond to those of c-Si and TiO$_{2}$, respectively. Colors correspond to different ratios of relative permeability with respect to relative permittivity. (b) Reflectance from a c-Si substrate nano-coated with a TiO$_{2}$ disk array for different symmetries in case of natural materials ($\mu_{r} = 1$): rectangular ($C_2$), square ($C_4$), and hexagonal ($C_6$) lattices. The lattice constants are $a_{x}$ = 625nm, $a_{y}$ = 400nm for $C_2$, $a$ = 500nm for $C_4$, and $a$ = 500nm for $C_6$.}
    \label{fig3}
\end{figure}

To that end, we now evaluate the response of a simplified system excluding ITO and a-Si films. This allows for a more straightforward assessment of the impact of the lattice symmetry by isolating it from other factors, like the optimization of the thicknesses of other layers. We consider square and hexagonal lattices of identical pitch for the disk arrays. First we resort to hypothetical materials. The electric permittivity of the substrate and disks correspond to those of c-Si and TiO$_{2}$, respectively. The magnetic permeabilities are varied from 60$\%$ to 90$\%$ of the value of the corresponding electric permittivities. Figure \ref{fig3}(a) shows that the hexagonal lattice produces a smaller reflection across the entire band for all the relative mismatches between $\epsilon_r$ and $\mu_r$. This happens even significantly away from the duality condition. As one may expect, the reflection grows with the mismatch.
The results in Fig. \ref{fig3}(a) motivate the study of hexagonal lattices in the case of natural materials with $\mu_r$=1. Figure \ref{fig3}(b) shows the reflectance from rectangular ($C_2$), square ($C_4$), and hexagonal ($C_6$) lattices. The $C_2$ array has been included for illustration purposes. In case of a rectangular array, we matched the unit cell area to the one of a square array, and for the hexagonal arrangement of the disks, the lattice constant was first taken to be the same as for the square array. For the square lattice, the cutoff wavelength at which nonzero diffraction orders vanish is equal to the array lattice constant $a$=500nm. In case of the regular hexagonal lattice, the wavelength should satisfy the condition $\lambda<0.5\sqrt{3}a$ to have nonzero diffraction orders. For a=500nm this yields 433nm. We observe that the $C_2$ lattice, which does not have enough rotational symmetry, is a less effective AR coating. We also observe that the higher symmetry of the hexagonal lattice improves the AR properties of the system compared to the square lattice. Instead of keeping the same lattice constant for both hexagonal and square lattices, we can also keep the ratio of the disks' surface area to the area of the lattice unit cell unchanged by increasing the lattice constant of the hexagonal lattice. Then, the hexagonal lattice still allows for more efficient suppression of reflection than the square array, but the difference is much smaller. This can be seen in Table \ref{tbl1} where all the considered disk arrangements are quantitatively compared through the integrated reflectance over the broad wavelength range (details in Methods). The results show the reduction of integrated values with increased rotational symmetry degree. 

\begin{table}
  \caption{Integrated reflectance for different arrangements of nano-particles}
  \label{tbl1}
  \begin{tabular}{  l | l | l | l | l } 
     & $C_2$ & $C_4$ & $C_6$ & $C_6$  \\
     & $a_{x}$ = 625nm, & $a$ = 500nm & $a$ = 500nm & $a$ = 537nm \\
    &  $a_{y}$ = 400nm  & & & \\
    \hline 
   Integrated reflectance (nm) & 161.01 & 144.37 & 118.24 & 141.77 \\ 
  \end{tabular}
  \end{table}

\section{Conclusions and outlook}
We have shown that helicity preservation and a high enough degree of discrete rotational symmetry are the relevant requirements for the design of nanoparticle arrays for reflection minimization in solar cells. For particle-lattice systems with high enough degree of discrete rotational symmetry ($C_{n>2}$), our numerical studies link the suppression of backscattering to the ability of the system to preserve the helicity (handedness) of the incident field. In an exemplary design, we have used TiO$_2$ disks placed on the front surface of a flat HJT solar cell stack in an electrically decoupled manner, and achieved a three-fold reduction of the current loss due to reflection with respect to a flat reference cell. We have also shown that a hexagonal lattice, with a higher degree of discrete rotational symmetry, can improve over the anti-reflection performance of a square lattice.

In light of the results contained in this article, solar cell performance can benefit from research directed towards the design of nanoparticles with broadband helicity preservation properties. In the case of disks, coated disks seem to be a plausible way to improve upon the helicity preservation properties of homogeneous disks. Such possibility is suggested by the fact that core-shell spherical designs improve upon the helicity preservation properties of homogeneous spheres \cite{Rahimzadegan2017}. Additionally, coated disks should also allow to enhance the response strength by aligning together other multipolar resonances beyond the dipoles \cite{Ruan2010}. The alignment of different resonances can then increase the helicity preserving bandwidth \cite{Abdelrahman2017,Abdelrahman2019}. 

\section{Methods} 
All the basic computations in the paper were done using the {\em JCMsuite} software.  

\subsection{Decomposition of scattered power into helicity components\label{sec:xixa}}
The decomposition of the scattered power into the power in each helicity component $\text{P}_{\pm}$ can be done in the following way.

For a plane wave of momentum $\mathbf{k}=[k_x,k_y,k_z]$, the polarization vectors corresponding to the $\pm 1$ helicity components are \cite[Sec. 2.2.4]{FerCorTHESIS}:
\begin{equation}
	\label{eq:hel}
	\hat{\mathbf{e}}_{\pm}(\mathbf{k})=\frac{1}{\sqrt{2}}[\hat{\mathbf{s}}(\mathbf{k})\pm\hat{\mathbf{p}}(\mathbf{k})], 
\end{equation}
where 
\begin{equation}
	\label{eq:sp}
	\begin{split}
		\hat{\mathbf{s}}(\mathbf{k})&=i\left[\sin(\phi)\hat{\mathbf{x}}-\cos(\phi)\hat{\mathbf{y}}\right],\\ 
		\hat{\mathbf{p}}(\mathbf{k})&=-\cos(\theta)cos(\phi)\hat{\mathbf{x}}-\cos(\theta)\sin(\phi)\hat{\mathbf{y}}+\sin(\theta)\hat{\mathbf{z}},
	\end{split}
  \end{equation}
and $\theta=\arccos(k_{z}/k)$, $\phi=\arctan(k_{y}/k_{z})$, with wave number $k = \sqrt {k_{x}^2+k_{y}^2+k_{z}^2}$. The $\hat{\mathbf{s}}(\mathbf{k})$ and $\hat{\mathbf{p}}(\mathbf{k})$ polarization vectors of Eq.~(\ref{eq:sp}) correspond essentially to the TE and TM polarizations, respectively. For example, for $\mathbf{k}=[0,0,k]$ we obtain $\hat{\mathbf{e}}_{\pm}(\mathbf{k})=\frac{1}{\sqrt{2}}(-i\hat{\mathbf{y}}\mp\hat{\mathbf{x}})$.

Each of the plane waves that compose the scattered field can be decomposed into its two helicity components as follows:
\begin{equation}
	\mathbf{E}(\mathbf{k})\exp(i\mathbf{k}\cdot\mathbf{r})=\left[\mathbf{E}_{+}+\mathbf{E}_{-}\right](\mathbf{k})\exp(i\mathbf{k}\cdot \mathbf{r}), 
\end{equation}
where
\begin{equation}
	\mathbf{E}_{\pm}(\mathbf{k})=\left\{\left[	\hat{\mathbf{e}}_{\pm}(\mathbf{k})\right]^\dagger \mathbf{E}(\mathbf{k})\right\}\hat{\mathbf{e}}_{\pm}(\mathbf{k})=E_{\pm}\hat{\mathbf{e}}_{\pm}(\mathbf{k}).
\end{equation}
Then, the normalized powers $\text{P}_{+}(\mathbf{k})$ and ${\text{P}_{-}(\mathbf{k})}$ corresponding to positive and negative helicity, respectively, can be computed as
\begin{equation}
	\text{P}_{\pm}(\mathbf{k})=\frac{|E_{\pm}|^2}{2Z_{0}\text{P}_\text{in}}, \label{eqn5}
\end{equation}

where the normalization is the power of the incident plane wave: $\text{P}_\text{in}=\frac{|\mathbf{E}_{0}|^2}{2}\sqrt{\epsilon_0/\mu_0} =\frac{|\mathbf{E}_{0}|^2}{2Z_0}$, where $\mathbf{E}_{0}$ is the amplitude and  $Z_{0}$ is the impedance of the medium. 

In Fig. \ref{fig1}(a), $P_\pm([0,0,-k])$ is shown, corresponding to the specular reflection direction. In Fig. \ref{fig2}(a), the powers reflected back to the air hemisphere are obtained through the integrals of $P_\pm(\mathbf{k})$ across the appropriate wave-vectors. The sum of the normalized powers of the two helicities shown in Fig. \ref{fig2}(a) is exactly equal to the reflectance computed using energy fluxes (see below), which constitutes a sanity check of our calculations.

\subsection{Reflectance}
The reflectance calculations are performed using the total EM energy flux $\Phi$ computed via integration of the EM energy flux density (Poynting vector) across layer interfaces as: 

\begin{equation}
  \text{R}=\frac{\mathrm{Re} \{ \Phi  \}}{\text{P}_\text{in}  \cdot \text{A}}, \label{eqn6}
\end{equation}
 where A is the unit cell area, and $\text{P}_\mathrm{in}\cdot \text{A}$  gives the EM power density. 

Integrated reflectance in nm was calculated as:

\begin{equation}
	\mathrm{IR}=\int_{\lambda_1}^{\lambda_2}\text{R}(\lambda) d\lambda, \label{eqn61}
\end{equation}
 
\subsection{Current loss per unit area}
The current loss due to reflection was calculated as:
\begin{equation}
	\ J_\mathrm{loss}=\int_{\lambda_1}^{\lambda_2}q_e\frac{SI(\lambda)\text{R}(\lambda)}{E_\mathrm{ph}(\lambda)} d\lambda, \label{eqn7}
\end{equation}
where $q_e$ is the electron charge, $SI(\lambda)$ is the spectral irradiance, and $E_\mathrm{ph}(\lambda)=hc/\lambda$ is the energy of a photon. For this calculation, air mass 1.5 global tilted irradiance raw data was taken from Ref. \onlinecite{gueymard1995smarts2}, and the simulated reflectance $\text{R}(\lambda)$ was interpolated accordingly.

\section{Acknowledgments}
We are grateful to the company JCMwave for their free provision of the FEM Maxwell solver JCMsuite, with which the simulations in this work have been performed. E.S. is pursuing her Ph.D. within the Karlsruhe School of Optics and Photonics (KSOP) and acknowledges financial support. This project has received funding from the European Union's Horizon 2020 research and innovation programme under the Marie Sk\l{}odowska-Curie grant agreement no. 675745. The authors also acknowledge support by KIT through the ``Virtual Materials Design'' (VIRTMAT) project by the Helmholtz Association via the Helmholtz program ``Science and Technology of Nanosystems'' (STN) and by the German Excellence Strategy through Deutsche Forschungsgemeinschaft (DFG) via the Excellence Cluster EXC 2082 ``3D Matter Made to Order'' (3DMM2O).

\bibliography{solar_cell_zbs}
\vbadness=99999
\end{document}